\documentclass[showpacs,aps,prl,twocolumn,superscriptaddress,amsmath,amssymb,nofootinbib]{revtex4}
\usepackage{graphicx}
\newcommand{\bra}[1]{\langle{#1}\rvert}
\newcommand{\ket}[1]{\lvert{#1}\rangle}
\newcommand{\braket}[1]{\mathinner{\langle{#1}\rangle}}
\newcommand{\proj}[1]{\mathinner{|{#1}\rangle\langle{#1}|}}
\newcommand{\abs}[1]{\lvert#1\rvert}

\newcommand{\cA}{\mathcal{A}}
\newcommand{\cB}{\mathcal{B}}
\newcommand{\cP}{\mathcal{P}} 
\newcommand{\cO}{\mathcal{O}}
\newcommand{\cS}{\mathcal{S}} 
\newcommand{\id}{\openone} 
\newcommand{\eb}[1]{\epsilon_{\bar{#1}}} 
\newcommand{\ef}[1]{\bar{\epsilon}_{#1}} 
\DeclareMathOperator{\tr}{Tr} 

\begin{document}

\title{Improved Error-Tradeoff and Error-Disturbance Relations}

\author{Xiao-Ming Lu}
\affiliation{Centre for Quantum Technologies, National University of Singapore, 3 Science Drive 2, Singapore 117543, Singapore}
\author{Sixia Yu} 
\affiliation{Centre for Quantum Technologies, National University of Singapore, 3 Science Drive 2, Singapore 117543, Singapore}
\author{Kazuo Fujikawa}
\affiliation{Centre for Quantum Technologies, National University of Singapore, 3 Science Drive 2, Singapore 117543, Singapore}
\affiliation{Mathematical Physics Laboratory, RIKEN Nishina Center, Wako 351-0198, Japan}
\author{C.H. Oh}
\affiliation{Centre for Quantum Technologies, National University of Singapore, 3 Science Drive 2, Singapore 117543, Singapore}
\affiliation{Physics department, National University of Singapore, 3 Science Drive 2, Singapore 117543, Singapore}

\begin{abstract}
Heisenberg's uncertainty principle is quantified by error-disturbance tradeoff relations, which have been tested experimentally in various scenarios.
Here we shall report improved new versions of various error-disturbance tradeoff relations by decomposing the measurement errors into two different components, namely, operator bias and fuzziness. 
Our improved uncertainty relations reveal the tradeoffs between these two components of errors, and imply various conditionally valid error-tradeoff relations for the unbiased and projective measurements. 
We also design a quantum circuit to measure the two components of the error and disturbance.
\end{abstract}

\maketitle


Uncertainty principle is one of the fundamental principles of quantum mechanics.
Heisenberg's original qualitative formulation is a tradeoff relation between the measurement precision of the position of a particle and the disturbance on its momentum~\cite{Heisenberg1927};
whilst the famous Kennard-Robertson uncertainty relation reflects the intrinsic quantum fluctuations within quantum states~\cite{Kennard1927,Robertson1929}. 
For unbiased measurements, Arthurs and Kelly, and Arthurs and Goodman revealed the extra uncertainty due to the measuring process~\cite{Arthurs1965, Arthurs1988}.
Only recently in 2003, Ozawa derived the first ``universally-valid'' error-disturbance uncertainty relation~\cite{Ozawa2003}. 
Subsequently, other uncertainty relations have been derived with reference to both the inherent fluctuation within the quantum states and the extra uncertainty from the measuring process~\cite{Ozawa2004_joint_measurement, Hall2004, Werner2004, Watanabe2011a, Watanabe2011b, Fujikawa2012, Branciard2013, DiLorenzo2013}.
Some of these uncertainty relations have been experimentally tested~\cite{Erhart2012, Rozema2012, Baek2013, Weston2013, Kaneda2013, Ringbauer2013}, using Ozawa's 3-state method~\cite{Ozawa2004} or Lund and Wiseman's weak measurement method~\cite{Lund2010}.

Most recently, operational definitions of state-independent measurement errors, which characterize the overall performance of measuring devices, have been proposed~\cite{Busch2013,Busch2013a,Buscemi2014} and a sort of operational issue was raised on Ozawa's formalism~\cite{Werner2004,Busch2004,Busch2013NoiseOperator,Korzekwa2013}. 
The basic fact behind is that the error in Ozawa's formalism in general cannot be determined only by comparing the outcomes of the approximate measurement and the ideal one.
Although this criticism from the point of view of operational significance is pertinent, those error quantities determined by the outcomes in the given state, which are henceforth simply referred to as {\em operational state-dependent errors}, cannot be used to formulate a state-dependent measurement uncertainty relation for all possible measurements with the lower bound being independent of the joint measurement.
For arbitrary given state and two ideal observables, the two corresponding operational state-dependent errors can simultaneously vanish for a particular choice of the joint measurement consisting of a legitimate quantum operation that {\em just} clones {\em this} specific state, followed by the two ideal measurements performed simultaneously on different copies.

Despite of the disputes on the appropriate definitions of the error and disturbance~\cite{Werner2004, DiLorenzo2013, Busch2013, Busch2007, Rozema2013, Ozawa2013, Dressel2013,BuschReply}, some theoretical efforts in the improvements for uncertainty relations include reviving Heisenberg's qualitative formulation by combining the intrinsic fluctuation and extra uncertainty~\cite{Fujikawa2012} or by invoking quantum estimation theory~\cite{Watanabe2011a,Watanabe2011b}, and tightening the inequality~\cite{Branciard2013}. 
In this Letter, we improve various error-tradeoff relations by formulating stronger inequalities in terms of two different components of the error and disturbance, namely the operator bias and fuzziness. 
This decomposition identifies an operationally significant part of Ozawa's definition of the total error, i.e., the fuzziness, based on which one can better understand the measurement uncertainty relation. 
Our improved uncertainty relations reflect not only the tradeoffs between the total errors but also the tradeoff between different components of errors. 
We provide a measurement circuit to experimentally measure these two components.

We first consider the joint measurements scenario and then proceed to the error-disturbance scenario.
According to Neumark's theorem, every generalized measurement can be implemented by a projective measurement on the system-plus-apparatus. 
Two noncommuting observables $A$ and $B$ can be approximately measured by the measurements of two commuting Hermitian operators $\cA$ and $\cB$ on the system-plus-apparatus~\cite{Arthurs1965}. Suppose that the system and apparatus is in a product state $\rho\otimes\rho_a$.
Following Ozawa's approach~\cite{Ozawa2003,Ozawa2004_joint_measurement}, the measurement error of $\cA$ can be measured by the root-mean-square (RMS) error 
\begin{multline}
(\epsilon_\cA)^2:=\braket{(\cA-A\otimes\id)^2}_{\rho\otimes\rho_a} \cr
=\braket{\overline{\cA^2}-\bar\cA^2}_\rho+\braket{(\bar\cA-A)^2}_\rho:={(\ef{\cA})}^2+{(\eb{\cA})}^2.
\end{multline}
Here the first line is exactly Ozawa's definition with $\id$ being the identity operator for the apparatus.
In the second line we defined a system observable $\bar{\cO}:= \tr_a[\cO(\id\otimes\rho_a)]$ for every observable $\cO$ of the whole system by the partial trace $\tr_a$ over the apparatus~\cite{Ozawa2003}. 
In this form the RMS error possesses the great advantage of being independent on the implementation of the generalized measurements on the system~\cite{Ozawa2004,Ozawa2005_operator_form} and can be regarded as a measure on the deviation of a generalized measurement from an ideal measurement. 
Suppose that $\sum_\alpha\alpha\cP_\alpha$ is the spectral decomposition of $\cA$ with distinct spectra $\alpha$, then $\{\bar\cP_\alpha\}$ is the positive-operator-valued measure (POVM) of the generalized measurement on the system. 
In the third line, we have decomposed the total error into components $\epsilon_{\bar\cA}$ and $\bar\epsilon_\cA$, called here as {\it operator bias} and {\it fuzziness}, respectively. This decomposition has also been considered by Hall~\cite{Hall2003} and Busch {\it et~al.}~\cite{Busch2004}, and by Di Lorenzo~\cite{DiLorenzo2013} for linear measurements~\cite{Braginsky1992}.

\paragraph{Operator bias $\eb{\cA}=\langle(\bar\cA-A)^2\rangle_\rho^{1/2}$.} We call $\eb{\cA}$ the {\em operator bias} because, firstly, the quantity $\eb{\cA}$ quantifies the difference between two observables $\bar\cA$ and $A$ in a given state, and the generalized measurement is unbiased when it is used to approximate the measurement of $\bar\cA$.
Secondly, this measure $\eb{\cA}$ bounds the expectation bias $\abs{\braket{\cA}_{\rho\otimes\rho_a}-\braket{A}_\rho}$ from above.
Moreover, $\eb{\cA}$ vanishes for all states of the system if and only if $\bar\cA=A$, i.e., Arthurs and Kelly's condition for unbiased measurement~\cite{Arthurs1965}.

\paragraph{Fuzziness $\ef{\cA}=\langle\overline{\cA^2}-\bar\cA^2\rangle_\rho^{1/2}$.}
The quantity $\ef{\cA}$ is well defined due to Kadison's inequality $\overline{\cA^2}\geq\bar\cA^2$~\cite{Kadison1952}, and is independent of the observable we intend to measure, i.e., $A$.
The fuzziness $\ef{\cA}$ vanishes for all states if and only if the measurement is a projective one.
Actually, any projective measurement of an observable $\cA$ on the entirety is an unbiased generalized measurement approximating the observable $\bar\cA$ on the system. 
In Ref.~\cite{Hall2003}, Hall interpreted $\ef{\cA}$ as the inherent {\em fuzziness} of generalized measurements because it is the smallest error between the given generalized measurement and all possible projective measurements. 
We note that for arbitrary operator $O$ and state $\rho$ of the system the following identity holds
\begin{equation}
\ef{\cA}^2 = \braket{(\cA-O\otimes\id)^2}_{\rho\otimes\rho_a}-\braket{(\bar\cA-O)^2}_\rho.
\end{equation}
For example we have $\ef{\cA}^2=\sigma_\cA^2-\sigma_{\bar\cA}^2$ when $O$ is taken as $\braket{\bar\cA}_\rho$ and $\ef{\cA}^2=\langle(\cA-\bar\cA\otimes\id)^2\rangle_{\rho\otimes\rho_a}$ when $O=\bar\cA$.

In a joint measurement of $A$ and $B$ by measuring two commuting observables $\cA$ and $\cB$, the Robertson uncertainty relation for the pair $\{\cA-\bar\cA\otimes\id,\cB-\bar\cB\otimes\id\}$ in the state $\rho\otimes\rho_a$ leads to a tradeoff of fuzziness 
\begin{equation}\label{fur}
\ef{\cA}\ef{\cB}\geq\frac12\abs{\braket{[\bar\cA,\bar\cB]}_\rho}.
\end{equation}
As an immediate result, together with three Robertson uncertainty relations for three pairs of observables $\{\bar\cA-A,B-\langle B\rangle_\rho\}$, $\{\bar\cB-B, A-\langle A\rangle_\rho\}$, and $\{\bar\cA-A,\bar\cB-B\}$, we obtain our main uncertainty relation
\begin{eqnarray}\label{ineq:new_type}
\ef{\cA}\ef{\cB}+\eb{\cA}\eb{\cB}+\eb{\cA}\sigma_B+\sigma_A\eb{\cB}\geq c_{AB}
\end{eqnarray}
with $c_{AB}=|\langle [A,B]\rangle_\rho|/2$ and $\sigma_A:=(\braket{A^2}_\rho-\braket{A}_\rho^2)^{1/2}$ is the standard deviation. Here the triangle inequality $|z_1+z_2|\le|z_1|+|z_2|$ for two arbitrary complex $z_1$ and $z_2$ has been used.

In 2004, Ozawa derived a universally-valid uncertainty relation for the joint measurements of two noncommuting observables as follows~\cite{Ozawa2004_joint_measurement}:
\begin{equation}\label{ineq:Ozawa_type}
\epsilon_\cA\epsilon_\cB+\epsilon_\cA\sigma_B+\sigma_A\epsilon_\cB\geq c_{AB}
\end{equation}
which has been experimentally tested. 
Ozawa's inequality Eq.(\ref{ineq:Ozawa_type}) can be derived from our inequality Eq.(\ref{ineq:new_type}), meaning that our inequality is stronger. 
To do so we have only to apply the Schwarz inequality $\epsilon_\cA \epsilon_\cB\geq\eb{\cA}\eb{\cB}+\ef{\cA}\ef{\cB}$ and the fact $\epsilon_\cA\geq\eb{\cA}$, to our inequality Eq.~(\ref{ineq:new_type}).
Hall also gave an alternative trade-off relation~\cite{Hall2004} $\epsilon_\cA\epsilon_\cB+\epsilon_\cA\sigma_\cB+\sigma_\cA\epsilon_\cB\geq c_{AB}$, which involves the standard deviation of $\cA$ instead of that of $A$.
Most recently, Weston {\it et~al.} derived another relation~\cite{Weston2013} $\epsilon_\cA(\sigma_\cB+\sigma_B)/2+\epsilon_\cB(\sigma_\cA+\sigma_A)/2\geq c_{AB}$.   
By using the same method leading to Eq.(\ref{ineq:new_type}), we obtain
\begin{equation}\label{ineq:Hall_improved}
\eb{\cA}\eb{\cB}+\ef{\cA}\ef{\cB}+\eb{\cA}\sigma_\cB+\sigma_\cA\eb{\cB}\geq c_{AB}, 
\end{equation}
which improves Hall's uncertainty relation because of the Schwarz inequality $\epsilon_\cA\epsilon_\cB\geq\eb{\cA}\eb{\cB}+\ef{\cA}\ef{\cB}$ and inequalities $\epsilon_\cA\geq\eb{\cA}$, $\epsilon_\cB\geq\eb{\cB}$, and an improved version of Weston {\it et al.}'s uncertainty relation
\begin{equation}\label{ineq:Weston_improved}
\ef{\cA}\ef{\cB}+\eb{\cA}\frac{\sigma_{\bar\cB}+\sigma_B}{2}+\eb{\cB}\frac{\sigma_{\bar\cA}+\sigma_A}{2}\geq c_{AB}.
\end{equation}

\begin{figure}
\begin{center} \includegraphics[width=7cm]{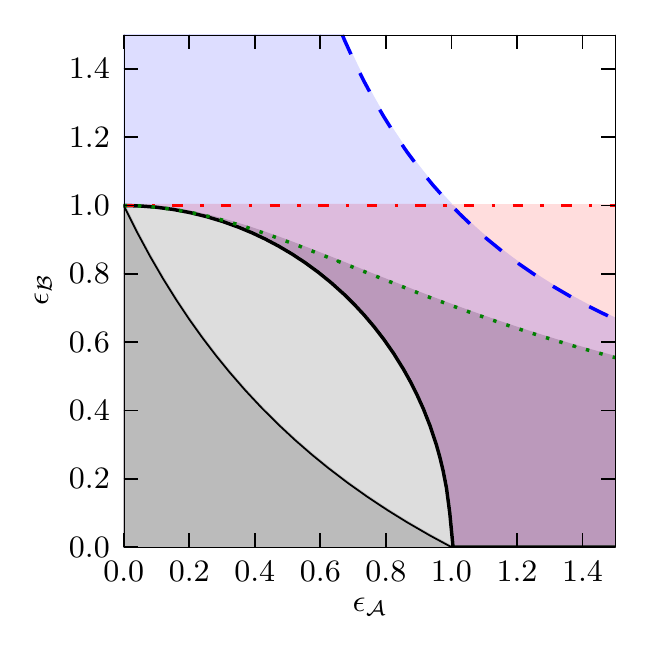} \end{center}
\caption{\label{fig:forbidden_region}
Forbidden regions (shaded) of errors $(\epsilon_\cA,\epsilon_\cB)$ implied by various error-tradeoff relations in the case of $\sigma_A=\sigma_B=c_{AB}=1$.
The region below the thin solid curve is forbidden by Ozawa's relation Eq.~(\ref{ineq:Ozawa_type}), while the region below the thick solid curve is forbidden by Branciard's tight (for pure states) error-tradeoff relation~\cite{Branciard2013}, which in this case reads $\epsilon_\cA^2+\epsilon_\cB^2\geq 1$.
The conditionally valid error-tradeoff relations, implied by our relation Eq.~(\ref{ineq:new_type}), forbid more errors $(\epsilon_\cA,\epsilon_\cB)$ for the cases of two unbiased measurements (dashed blue curve), unbiased for $A$ and projective for $B$ (dash-dotted red line), and unbiased for $A$ and unconditional for $B$ (dotted green curve).}
\end{figure}

Several conditionally valid error-tradeoff relations in terms of total errors can be derived from our relation Eq.(\ref{ineq:new_type}):
i)  When the measurement for $A$ is unbiased, our relation implies $\sqrt{\epsilon_\cA^2+\sigma_A^2}\epsilon_\cB\geq\ef{\cA}\ef{\cB}+\sigma_A\eb{\cB}\geq c_{AB}$. 
ii) When the measurement for $A$ is unbiased and that for $B$ is projective, our relation becomes $\epsilon_{\cB}\sigma_A\geq c_{AB}$. 
iii) When the two measurements are unbiased, our relation becomes $\epsilon_{\cA}\epsilon_{\cB}\geq c_{AB}$, which is the error-tradeoff version of Arthurs and Kelly's relation~\cite{Arthurs1965,Arthurs1988,Appleby1998}. 
Because the projective measurements are common and unbiased measurements are desired as good measurements, the above conditionally valid error-tradeoff relations are of practical interest.
As shown in Fig.~\ref{fig:forbidden_region}, these conditionally valid error-tradeoff relations will forbid more combinations of the total errors.

Our relation Eq.(\ref{ineq:new_type}) also reflects tradeoffs between operator bias and fuzziness. 
Particularly, Arthurs and Kelly's relation states only that the unbiasedness must cause the fuzziness, while our relation also expresses that the sharpness must cause the operator bias.
This tradeoff between operator bias and fuzziness can be used to explicitly reveal the inconsistency of some settings of the measuring processes, e.g., the algebraic inconsistency between the unbiased disturbance assumption and the occurrence of a zero-error of the measurement in a state~\cite{Fujikawa2013}.

The fuzziness itself is operationally significant: its square equals the added variance of $\cA$ compared with that of $\bar\cA$, for which the measurement device is unbiased.
Equation (\ref{fur}) is the error-tradeoff relation for unbiased measurements derived by Appleby~\cite{Appleby1998}, when $\bar\cA=A$ and $\bar\cB=B$ are assumed. 
Here, Eq.~(\ref{fur}) is interpreted in another way as a fuzziness-tradeoff for a given joint measurement device. 
We may have neither the knowledge about the operators $\bar\cA$ and $\bar\cB$ nor the measurement devices performing projective measurement of them; we just assume that we have the joint measurement device.
In such a circumstance, the lower bound in Eq.~(\ref{fur}) can still be experimentally obtained, which will be discussed later.

Other error-tradeoffs are possible. As an application of the fuzziness uncertainty relation Eq.(\ref{fur}), together with Robertson's uncertainty relation $\sigma_{\bar\cA}\sigma_{\bar\cB}\geq\abs{\braket{[\bar\cA,\bar\cB]}_\rho}/2$, and the Schwarz inequality $\sigma_\cA\sigma_\cB\geq\ef{\cA}\ef{\cB}+\sigma_{\bar\cA}\sigma_{\bar\cB}$, one can derive a Robertson-like uncertainty relation $\sigma_\cA\sigma_\cB\geq\abs{\braket{[\bar\cA,\bar\cB]}_\rho}$ for a pair of commuting observables.
Recall that a naive application of Robertson's relation for commuting observables $\cA$ and $\cB$ gives rise to only a trivial bound $\sigma_\cA\sigma_\cB\geq 0$.
The emergence of a nontrivial lower bound can be traced back to the fact that all the expectation values are taken over a product state $\rho\otimes\rho_a$, which is imposed by the measuring process. For other examples, by  using the same arguments leading to Eq.(\ref{ineq:new_type}), one can derive  the following two error tradeoffs
\begin{eqnarray}
\sigma_{\bar{\cA}}\sigma_{\bar{\cB}}+\eb{\cA}\eb{\cB}+\eb{\cA}\sigma_B+\sigma_A\eb{\cB}\geq c_{AB},\label{robertson_type}\\
(\eb{\cA}+\sigma_{\bar{\cA}})(\eb{\cB}+\sigma_{\bar{\cB}})\geq c_{AB},\label{heisenberg_type}	
\end{eqnarray}
in which $[\cA,\cB]=0$ is implicitly contained in the definitions of $\bar{\cA}$ and $\bar{\cB}$.
The relations  Eq.(\ref{ineq:new_type}) and Eq.(\ref{robertson_type}) are closely related but independent. 

\begin{figure}
\includegraphics[width=8.6cm]{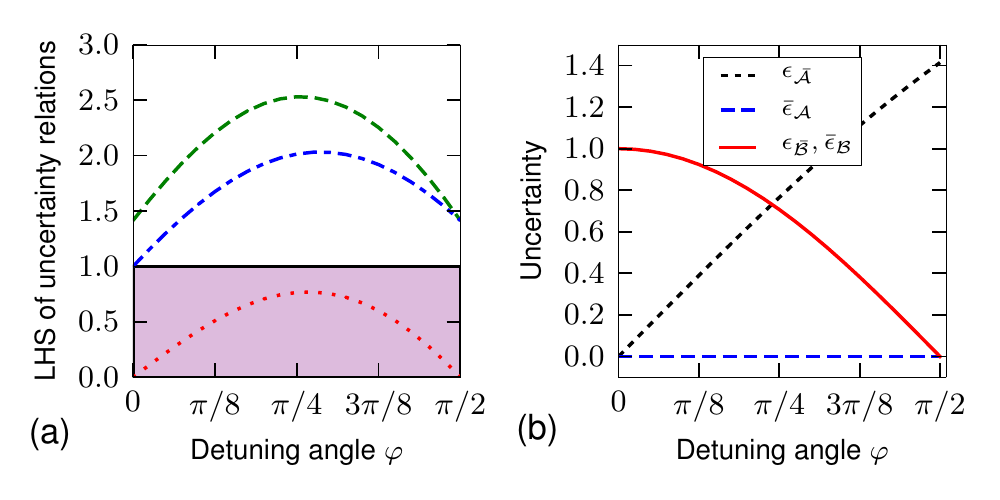}
\caption{\label{fig:Vienna_experiment}
Error-disturbance uncertainty relations in the case of the experiment in Ref~\cite{Erhart2012}.
In the Fig (a), the left hand sides (LHS) of Ozawa's relation (dashed green curve) and that of our relation~(\ref{ineq:new_type}) (dash-dotted blue curve) as well as the product of the error and disturbance (dotted red curve) are plotted.
The red shading indicates the region where these relations are violated. 
The components of the error and disturbance are given by in figure (b).
Here, the observables of interest are $A=X$ and $B=Y$, the state of the spin is the eigenstate of $Z$ with eigenvalue $+1$, where $X$, $Y$, and $Z$ are the Pauli matrices.
The measurement performed for $A$ is the projective measurement of $X_\varphi:=\cos\varphi X+\sin\varphi Y$.}
\end{figure}

\paragraph{Error-disturbance scenario.}
In the error-disturbance scenario, one considers the tradeoff between the error of the measurement for $A$ and the caused disturbance on another observable $B$. 
The disturbance is naturally measured by the deviation between the ideal measurement of $B$ in the original state and that in the post-measurement states after the measurement for $A$.
This can be formally recast into the scenario of joint measurements by combining the two successive measuring apparatus as a new apparatus to perform joint measurements for $A$ and $B$, and measuring the disturbance by the error of joint measurement for $B$~\cite{Ozawa2003}. 
A comparison between our relation (\ref{ineq:new_type}) and Ozawa's error-disturbance uncertainty relation in the Vienna experiment~\cite{Erhart2012}
is shown in Fig.~\ref{fig:Vienna_experiment}. 
Ozawa's inequality cannot be saturated for any detuning angle for the measurement, while our inequality is saturated when the measurement for $X$ is precise, i.e., at $\varphi=0$. 

\begin{figure}
\begin{center}\includegraphics[width=7cm]{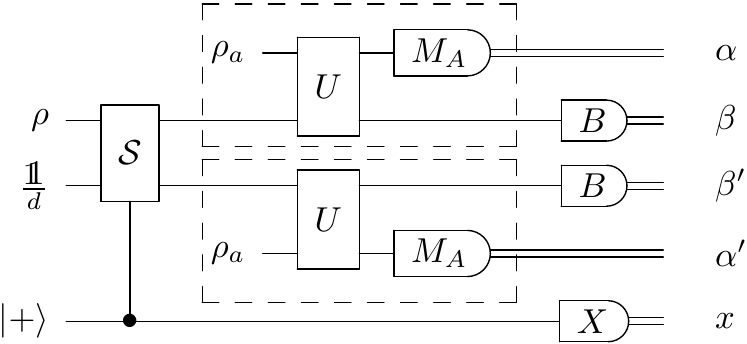}\end{center}
\caption{\label{fig:quantum_circuit}
Quantum circuit to measure the fuzziness $\ef{\cA}$ and $\ef{\cB}$.
The dashed boxes stand for a generalized measurement of $A$, which is without loss of generality implemented by the indirect measurement model.}
\end{figure}

{\em Experimental feasibility.}
To experimentally measure the operator bias and the fuzziness, we need to measure the expectation value $\braket{\bar{\mathcal{A}}^2}_\rho$.
In principle $\braket{\bar{\mathcal{A}}^2}$ can be obtained by the projective measurement of $\bar\cA$, which is not accessible in the recent experiments~\cite{Erhart2012,Rozema2012,Baek2013,Weston2013,Kaneda2013,Ringbauer2013}.
For finite-dimensional systems, $\braket{\bar{\mathcal{A}}^2}$ can be experimentally measured, provided that we have two apparatus implementing the same measurement. 
Suppose that we have an auxiliary qubit and an reference system whose Hilbert space has the same dimension $d$ as that of the system to be measured. 
Initially the system is prepared in the given state $\rho$, the reference system is prepared in the maximal mixed state $\id/d$, and the auxiliary qubit is prepared in the +1 eigenstate $\ket{+}$ of $X=\ket{0}\bra{1}+\ket{1}\bra{0}$. 
After the preparation, a controlled swap 
is performed with the qubit as the control and the system and reference system as the target, see Fig~\ref{fig:quantum_circuit}.

The quantum circuit measuring the fuzziness is schematically drawn in Fig.~\ref{fig:quantum_circuit}.
We perform a joint measurement on the system and the reference system together with an $X$ measurement on the qubit with outcomes labeled with random variables $\alpha,\beta,\alpha^\prime,\beta^\prime,$ and $x$ respectively. The experimental data are nothing else than a joint probability distribution of these five random variables and let $\mathbb{E}[\cdot]$ denote the average of the outcomes.
It turns out (see Supplemental Material~\cite{SM})
\begin{eqnarray}\label{mf}
\bar\epsilon_\cA=\sqrt{d{\mathbb{E}[\alpha(\alpha-\alpha^\prime)x]}}, \;
\bar\epsilon_\cB=\sqrt{d{\mathbb{E}[\beta(\beta-\beta^\prime)x]}}.
\end{eqnarray}
Interestingly, by replacing the measurement of $X$ on the auxiliary qubit by that of $Y$, the lower bound of the relation Eq.~(\ref{fur}) can be obtained as $\frac{1}{2}|\langle\bar{\mathcal{A}},\bar{\mathcal{B}}\rangle_{\rho}|=d|\mathbb{E}(\alpha\beta^{\prime}y)|$ with $y$ being the random variable for the outcomes of $Y$ (see Supplemental Material~\cite{SM}). Thus, the fuzziness-tradeoff Eq.~(\ref{fur}) itself can be experimentally tested.

To experimentally measure the total error (disturbance), we have only to replace the joint measurement on the reference system by an ideal measurement of $A$ or $B$.
If we denote the outcomes of the ideal measurement of $A,B$ by random variables $a,b$, then the total error for $A$ and the disturbance on $B$ are (see Supplemental Material~\cite{SM})
\begin{eqnarray}\label{me}
\epsilon_{\cA}=\sqrt{d{\mathbb{E}_A[(\alpha-a)^2x]}},\;
\epsilon_{\cB}=\sqrt{d{\mathbb{E}_B[(\beta-b)^2x]}}.
\end{eqnarray}
The operator bias can be evaluated via the total error, the total disturbance and their fuzziness.
In the case of $d=2$, the controlled swap together with the $X$ measurement can be replaced by a projection to the singlet state  of the system and the reference system (see Supplemental Material~\cite{SM}).

\paragraph{Conclusions and discussions.}
In summary, we have improved various error-tradeoff relations of experimental interest by decomposing the RMS error for generalized measurement into two components.
Our uncertainty relations not only tighten previous results, but also explicitly reveal the tradeoff relation between these two different types of the measurement errors. Also we obtain various conditionally valid error-tradeoff relations, including the Arthur-Kelly-Goodman relation as a special case. 
Like Ozawa's relation, our improved version Eq.(\ref{ineq:new_type}) is also experimentally testable, as demonstrated by a quantum circuit to measure the fuzziness.

It is difficult to define a fully satisfying state-dependent measurement error and formulate the underlying tradeoff.
At the cost of missing a universal operational significance, Ozawa~\cite{Ozawa2003} formulated his relation for arbitrary state and joint measurement in terms of the simple RMS error with certain advantages.
From the perspective of the state-operator duality, this RMS error can be practically used to investigate the inaccuracy of a general class of measurements devices {\em through} a fixed state. 
Nevertheless, one should keep in mind the issue of operational significance when interpreting it as the measurement error {\em with respect to} the fixed state.
Some aspects of this issue can be avoided by the choice of the class of measurement devices.
For example, in the Vienna experiment~\cite{Erhart2012}, 
the outcome distributions of $X_\varphi$ are the same as those of the ideal one, whereas the RMS error is not zero.
However, if we consider a more general class of measurements, those of $\sin\theta\cos\varphi X+\sin\theta\sin\varphi Y+\cos\theta Z$, only a measure-zero subset causes this difficulty.
In fact, 
the measurements producing the same outcome distribution in the fixed state can be considered as an equivalent class.
Only for a class of measurements such that any two of them are in different equivalent classes, e.g., $\sin\theta X+\cos\theta Z$ in the Vienna experiment, the RMS error could be viewed to be consistent with the operational processes of comparing the outcome distributions.

Finally we briefly compare our work with the recent profound paper~\cite{Busch2013}. 
Busch, Lahti, and Werner proposed an operational state-independent measurement error through the Wasserstein-2 metric on outcome distributions and formulated the Heisenberg-type relation~\cite{Busch2013}; whilst our entire analysis is based on Robertson's relation, namely, the specification of not only initial states but also measurements is based on standard deviations.  
Besides, Buscemi {\it et~al.} proved an entropic state-independent the error-disturbance relation~\cite{Buscemi2014}, and Coles and Furrer proved an entropic state-independent ``residual ignorance''-disturbance relation~\cite{Coles2013}.
These operationally significant approaches should be kept in mind in future works on measurement uncertainty relations.

\begin{acknowledgments}
Lu thanks Patrick Coles for helpful discussions and Antonio Di Lorenzo for helpful communications.
This work is supported by National Research Foundation and Ministry of Education, Singapore (Grant No. WBS: R-710-000-008-271), and also funded by the Singapore Ministry of Education (partly through the Academic Research Fund Tier 3 MOE2012-T3-1-009).    
\end{acknowledgments}

\newpage
\setcounter{equation}{0}
\renewcommand{\theequation}{S\arabic{equation}}
\begin{center}
{\bf Supplementary Material}
\end{center}

At first we shall present the detailed derivation of our uncertainty relations. 
The Robertson uncertainty relation for a pair of observables $\{O_1,O_2\}$ in the state $\rho$ refers to
the following inequality
\begin{equation}
\sqrt{\langle O_1^2\rangle_\rho}\sqrt{\langle O_2^2\rangle_\rho}\ge\frac 12|\langle[O_1,O_2]\rangle_\rho|.
\end{equation}
As an example, uncertainty relation Eq.(4) in the main text holds because
\begin{eqnarray}
&&\ef{\cA} \ef{\cB} + \eb{\cA} \eb{\cB} + \eb{\cA} \sigma_B + \sigma_A \eb{\cB} \nonumber\\ 
&\geq&\frac12|\langle[\bar \cA,\bar\cB]\rangle_\rho|+\frac12|\langle[\bar\cA-A,\bar\cB-B]\rangle_\rho|\nonumber\\
&&+\frac12|\langle[\bar\cA-A,B]\rangle_\rho|+\frac12|\langle[A,\bar\cB-B]\rangle_\rho|\nonumber\\
&\geq& \frac12|\langle[\bar \cB,\bar\cA]+[\bar\cA-A,\bar\cB]+[A,\bar\cB-B]\rangle_\rho|\nonumber\\ &=& c_{AB}
\end{eqnarray}
and the uncertainty relation Eq.~(7) in the main text holds because
\begin{eqnarray}
    &&\ef{\cA}\ef{\cB}+\eb{\cA} \frac{\sigma_{\bar\cB} + \sigma_B}{2} + \eb{\cB} \frac{\sigma_{\bar\cA} + \sigma_A}{2} \nonumber\\
    &\geq&\frac12|\langle[\bar\cA,\bar\cB]\rangle_\rho|\nonumber\\ 
    &&+\frac14|\langle[\bar\cA-A,\bar\cB]\rangle_\rho|+\frac14|\langle[\bar\cA-A,B]\rangle_\rho|\nonumber\\
    &&+\frac14|\langle[\bar\cA,\bar\cB-B]\rangle_\rho|+\frac14|\langle[A,\bar\cB-B]\rangle_\rho|\nonumber\\
    &\geq&\frac12|\langle[\bar\cA,\bar\cB]\rangle_\rho|+\frac12|\langle[A,B]-[\bar \cA,\bar\cB]\rangle_\rho|\nonumber\\
    &\geq& c_{AB}.
\end{eqnarray}

Next we shall show that the fuzziness and total error can be read out directly from the output of our quantum circuit, i.e., Eq.~(10) and Eq.~(11) in the main text. The Hilbert space of the whole system, including the system, the reference system, and a qubit, is $\mathbb{C}^d \otimes \mathbb{C}^d \otimes \mathbb{C}^2$. Initially the whole system is prepared in a product state $\rho_i = \rho \otimes \frac{\openone}{d} \otimes \proj{+}$, i.e., the system is prepared in the given state $\rho$, 
the reference system is prepared in the maximal mixed state $\id/d$, and the auxiliary qubit is prepared in the +1 eigenstate $\ket{+}$ of $X = \ket{0}\bra{1}+\ket{1}\bra{0}$.
After the controlled swap operation $$U_\mathrm{cs}=\id \otimes \id \otimes \proj{0} + \cS \otimes \proj{1}$$ with $\cS$ being the swap operator and the qubit as control, the state of the whole system is given by
\begin{eqnarray}
    \rho_f  &= & U_\mathrm{cs} \rho_i U_\mathrm{cs}^\dagger \nonumber\\
            &= & \frac{1}{2d}\left[\rho \otimes {\openone} \otimes \proj{0} + {\openone} \otimes \rho \otimes \proj{1} \right.\nonumber\\ 
            & &\left. + (\rho \otimes {\id})\cS \otimes \ket{0}\bra{1} + \cS(\rho \otimes {\id}) \otimes \ket{1}\bra{0}\right].
\end{eqnarray}
Let $\{|\beta\rangle\langle\beta|\},\{|\alpha\rangle\langle\alpha|\}$ be the projections corresponding to the measurements made on the system and its apparatus, respectively, and $\{|\beta^\prime\rangle\langle\beta^\prime|\},\{|\alpha^\prime\rangle\langle\alpha^\prime|\}$ be the projections corresponding to the measurements made on the reference system and its apparatus, respectively. Also we denote by $\{|x\rangle\langle x|\}$  the projections corresponding to the measurements made on the qubit where $|x\rangle$ are two eigenstates of $X$ with $x=\pm$. The corresponding generalized measurements on the system are described by the following POVM
$$M_{\alpha\beta}=\tr_{a}[\rho_aU^\dagger(|\beta\rangle\langle\beta|\otimes|\alpha\rangle\langle\alpha|)U]$$ while the corresponding POVM on the reference system is given by $M_{\alpha^\prime\beta^\prime}$. Thus the probability distribution obtained at the output of our circuit reads
\begin{eqnarray}\label{pd}
P(\alpha,\beta,\alpha^\prime,\beta^\prime,x)=\tr[(M_{\alpha\beta} \otimes M_{\alpha^\prime\beta^\prime} \otimes |x\rangle\langle x|) \rho_f].
\end{eqnarray}
As a result we have
\begin{eqnarray}
\mathbb{E}[\alpha\alpha^\prime x]&=&\sum \alpha\alpha^\prime x P(\alpha,\beta,\alpha^\prime,\beta^\prime,x)\nonumber\\
&=&\tr[(\bar\cA \otimes \bar\cA \otimes X) \rho_f]\nonumber\\
&=&\frac1d\langle \bar\cA^2\rangle_\rho
\end{eqnarray}
and
\begin{eqnarray}
\mathbb{E}[\alpha^2x]&=&\sum \alpha^2x P(\alpha,\beta,\alpha^\prime,\beta^\prime,x)\nonumber\\
&=&\tr[(\overline{\cA ^2} \otimes \id \otimes X) \rho_f]\nonumber\\
&=&\frac1d\langle \overline{\cA ^2}\rangle_\rho
\end{eqnarray}
from which Eq.~(10) in the main text follows. Here we have used the definition $\bar O=\tr_a[\rho_a U^\dagger( \id\otimes O)U]$ and the following identity
\begin{align}
    &\tr[(O_1 \otimes O_2 \otimes X) \rho_f] \nonumber\\
    =& \frac{1}{2d}\tr\left[(O_1 \otimes O_2) (\rho \otimes {\id})\cS  + (O_1 \otimes O_2) \cS(\rho \otimes {\id}) \right] \nonumber\\
    =& \frac{1}{2d}\tr(O_1 \rho O_2 + \rho O_1 O_2)\nonumber\\
    =& \frac{1}{2d}\langle O_1 O_2 + O_2 O_1 \rangle_\rho,
\end{align}
for any operators $O_1$ and $O_2$ on the system and reference system, respectively, in which the property of the swap operator $\tr[(O_1 \otimes O_2)\cS]=\tr(O_1 O_2)$ for any operator $O_1$ and $O_2$ has been used. 

\begin{figure}[tb]
\begin{center}
\raisebox{2cm}{(a)}\hskip10pt        
\includegraphics[width=6.5cm]{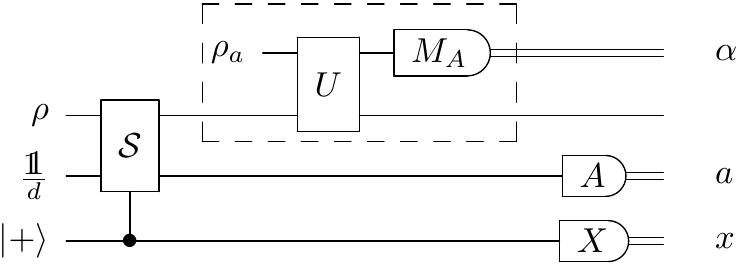}\\
\vspace{.5cm}
\raisebox{2cm}{(b)}\hskip10pt
\includegraphics[width=6.5cm]{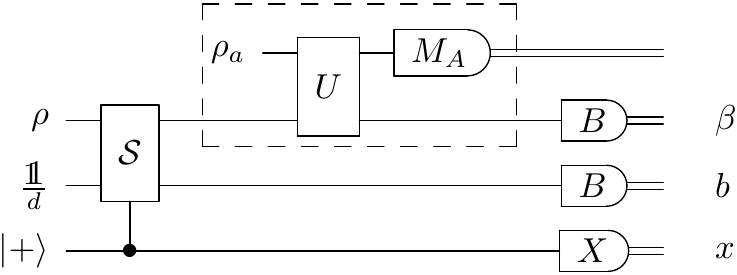}\\
\vspace{.5cm}
\raisebox{3cm}{(c)}\hskip10pt
\includegraphics[width=6.5cm]{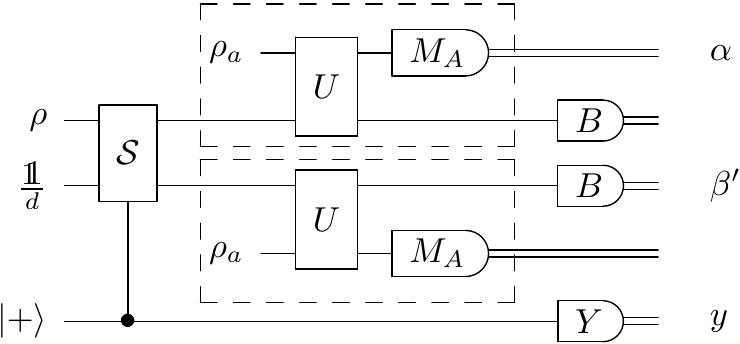}
\end{center}
\caption{\label{fig:measure_total}Quantum circuits to measure (a) the total error $\epsilon_\cA$,  (b) total disturbance $\eta_\cB$, and (c) $|\langle [\bar\cA,\bar\cB]\rangle_\rho|$.
The dashed boxes stand for an indirect measurement model implemented for the generalized measurement that approximates the measurement of $A$.}
\end{figure}

Similarly if the unitary evolution of the reference system and its apparatus is switched off and a projective measurement of $A$ is made directly on the reference system, see Fig.~\ref{fig:measure_total} (a), we obtain at the output the following probability distribution
\begin{eqnarray*}
P_A(\alpha,\beta,a,x)&=&\tr[(M_{\alpha\beta} \otimes |a\rangle\langle a| \otimes |x\rangle\langle x|) \rho_f],
\end{eqnarray*}
where ${|a\rangle}$ are eigenstates of $A$. As a result we obtain
\begin{eqnarray}
    {\mathbb{E}_A[(\alpha-a)^2x]}&=&\sum (\alpha-a)^2x P_A(\alpha,\beta,a,x)\nonumber\\
    &=&\tr[(\overline{\cA^2} \otimes \id - 2 \bar\cA \otimes A  + \id\otimes A^2) \otimes X \rho_f]\nonumber\\
    &=&\frac 1d \langle \overline{\cA^2}-A\bar\cA - \bar\cA A +A^2 \rangle_\rho = \frac 1d \epsilon_\cA^2.
\end{eqnarray}
Similarly, if we make a projective measurement of $B$ instead of $A$ on the reference system, see Fig.~\ref{fig:measure_total} (b), we obtain the relation for the disturbance as $$\epsilon_{\cB}=\sqrt{d{\mathbb{E}_B[(\beta-b)^2x]}}.$$
Also, the expectation value of the commutator $[\bar\cA,\bar\cB]$ can be experimentally measured by the circuit in Fig.~\ref{fig:measure_total} (c), as $$\mathbb{E}_C(\alpha\beta^{\prime}y) = \tr[(\bar\cA\otimes\bar\cB\otimes Y)\rho_f] = \frac{-i}{2d}\langle[\bar\cA,\bar\cB]\rangle_{\rho}.$$

Suppose that the system and the reference qubit are prepared in the state $\rho\otimes\id/d$.
Instead of a controlled swap, we make a measurement $\{P_{as},\id\otimes\id-P_{as}\}$  on the whole system and let $x$ be the outcome taking value $-1$ if the effect $P_{as}$ occurs and $+1$ otherwise. 
Here $$P_\mathrm{as}:=\frac{\id\otimes\id-\cS}2$$ is the projector  onto the antisymmetric subspace of the bipartite Hilbert space of the system and the reference system. In the case of $d=2$ it becomes  the projector onto the singlet state $\ket{\Psi}=(|01\rangle-|10\rangle)/\sqrt2$. 
And then we make joint measurements $M_{\alpha\beta}$ and $M_{\alpha^\prime\beta^\prime}$ of $A$ and $B$ on both the system and the reference system. The output probability distribution is identical with Eq.(\ref{pd}) and therefore the fuzziness can be obtained directly. The advantage of this approach lies in the fact that for two qubits a test of whether they are in a singlet state or not, e.g., by Bell-state measurements, is more accessible than a 3-qubit controlled swap  gate.

\end{document}